\documentstyle[12pt]{article}
\newcommand{\journal}[4]{{\em #1~}#2\,(19#3)\,#4;}

\newcommand{\pr}{\journal {Phys. Rev.}}

\newcommand{\jmp}{\journal {J. Math. Phys.}}

\newcommand{\cmp}{\journal {Comm. Math. Phys.}}

\newcommand{\np}{\journal {Nucl. Phys.}}
\newcommand{\pl}{\journal {Phys. Lett.}}

\newcommand{\prep}{\journal {Phys. Reports}}

\newcommand{\annp}{\journal {Ann. Phys. (N.Y.)}}

\makeatletter
\@addtoreset{equation}{section}
\makeatother

\def\Lp{\displaystyle{\biggl(}}
\def\Rp{\displaystyle{\biggr)}}
\def\LP{\displaystyle{\Biggl(}}
\def\RP{\displaystyle{\Biggr)}}
\newcommand{\lp}{\left(}\newcommand{\rp}{\right)}

 \renewcommand{\S}{\Sigma}

\newcommand{\complex}{{\kern .1em {\raise .47ex
\hbox {$\scriptscriptstyle |$}}
    \kern -.4em {\rm C}}}
\newcommand{\real}{{{\rm I} \kern -.19em {\rm R}}}
\newcommand{\rational}{{\kern .1em {\raise .47ex
\hbox{$\scripscriptstyle |$}}
    \kern -.35em {\rm Q}}}
\renewcommand{\natural}{{\vrule height 1.6ex width
.05em depth 0ex \kern -.35em {\rm N}}}

\newcommand{\pad}[2]{{\frac{\partial #1}{\partial #2}}}
\newcommand{\fud}[2]  {{\displaystyle{\frac{\delta #1}{\delta #2}}}}

\newcommand{\sla}{\raise.15ex\hbox{$/$}\kern -.57em}

\newcommand{\twiddle}{\lower.9ex\rlap{$\kern -.1em\scriptstyle\sim$}}

\newcommand{\equ}[1]{(\ref{#1})}

\newcommand{\eq}{\begin{equation}}
\newcommand{\eqn}[1]{\label{#1}\end{equation}}
\newcommand{\eea}{\end{eqnarray}}
\newcommand{\eqa}{\begin{eqnarray}}
\newcommand{\eqan}[1]{\label{#1}\end{eqnarray}}
\newcommand{\ba}{\begin{array}}
\newcommand{\ea}{\end{array}}
\newcommand{\eqac}{\begin{equation}\begin{array}{rcl}}
\newcommand{\eqacn}[1]{\end{array}\label{#1}\end{equation}}

\renewcommand{\pad}[2]{{\displaystyle{\frac{\partial #1}{\partial #2}}}}

\newcommand{\intx}{\int d^4 \! x \, }

\begin{document}
\def\ftoday{{\sl  \number\day \space\ifcase\month
\or Janvier\or F\'evrier\or Mars\or avril\or Mai
\or Juin\or Juillet\or Ao\^ut\or Septembre\or Octobre
\or Novembre \or D\'ecembre\fi
\space  \number\year}}


{\large 
\titlepage

\begin{center}

{\huge Perturbation Theory for Antisymmetric Tensor Fields in Four Dimensions}

\vspace{1cm}

{\Large N. Maggiore}

\vspace{.5cm}

{\it
Dipartimento di Fisica -- Universit\`a di Genova\\
Istituto Nazionale di Fisica Nucleare -- sez. di Genova\\
Via Dodecaneso, 33 -- 16146 Genova (Italy)}

\vspace{1cm}
and

\vspace{.4cm}

{\Large S. P. Sorella}\footnote{Supported in part by the
Swiss National Science Foundation}

\vspace{.5cm}
{\it D\'epartement de Physique Th\'eorique, Universit\'e de Gen\`eve\\
CH--1211 Gen\`eve 4, Switzerland}

\end{center}

\vspace{1.7cm}

\begin{center}
\bf ABSTRACT
\end{center}
{\it
Perturbation theory for a class of topological field theories
containing antisymmetric tensor fields is considered.
These models are characterized by
a supersymmetric structure which allows to establish their perturbative
finiteness.
}

\vfill
GEF-Th-6/1992 \\
UGVA--DPT 1992/04-761 \hfill April 1992
\newpage

\section{Introduction}

The antisymmetric tensor fields, or $BF$-models, were introduced many
years ago in connection with string theories and nonlinear sigma
models~\cite{ogiev}.

More recently they have been the object of a renewed interest due to their
topological nature~\cite{birm}. Antisymmetric tensor fields can be
used, for instance, to compute topological invariants~\cite{horo} which
generalize the three-dimensional linking number~\cite{perish}.

These models are also studied for their connection with
lower
dimensional quantum gravity; in particular, the Einstein-Hilbert gravity
in three space-time dimensions, with or without cosmological constant, can be
naturally formulated in terms of the $BF$-models~\cite{birm,witten,eno}.

{}From a pure field theoretical point of view, the $BF$-models are known to
require a highly non trivial quantization~\cite{grisaru,gms,abud}
due to the presence
of zero modes, which implies several ghost generations for the gauge
fixing procedure.
Moreover, as shown for the three-dimensional case~\cite{ms}, they are
expected to be an example of finite theories.

The purpose of this work is to give a proof of the perturbative finiteness for
the more complex four dimensional case, relying on the existence of a
supersymmetric structure found by the authors~\cite{gms}.
This structure, whose topological origin is manifest when adopting a Landau
gauge,
is a common feature of a large class of topological
models~\cite{ms,chern}.

Although the choice of an
axial-type gauge, which trivializes the ghost sector, appears to be
more convenient than
the covariant Landau gauge, the adoption of the latter turns out
to be very useful
to discuss the higher dimensional generalization of the model~\cite{olivier}.
Indeed in dimensions higher than four, the $BF$-models naturally
contain the generalized Chern-Simons terms~\cite{zumino}
which, being not trivial
even
with a noncovariant gauge choice, are easily handled in the
Landau gauge, thanks to the existence of the above mentioned supersymmetric
structure.

The essence of the method is to encode all the constraints defining the model
(BRS invariance, supersymmetry, ghost equations, $\ldots$) into an extended
BRS operator by the introduction of new global ghosts.
The quantized theory is then controlled by a generalized Slavnov identity
which contains all the symmetries. This technique turns out to be
very powerful when dealing with an algebraic structure described by several
operators. In particular, the search of the anomalies for each single operator
is reduced to a unique cohomology problem for the extended Slavnov
operator.

The paper is organized as follows: sect.2 contains a brief review of the
algebraic classical properties. In sect.3 we discuss the absence of local
counterterms and, finally, in sect.4 we characterize the anomalies.

\section{The classical model and the algebraic structure}

This section is devoted to a brief summary of the classical properties of the
model.

Following~\cite{gms}, the model is characterized by a complete
gauge--fixed classical action
\eq
\Sigma=S_{inv}+S_{gf}+S_{ext}\ ,
\eqn{actionclass}
where
\eq
S_{inv} = - {1 \over 4}
  \intx {\ } \varepsilon^{\mu\nu\rho\sigma}
                     F^{a}_{\mu\nu} B^{a}_{\rho\sigma}
\eqn{actioinv}
is the topological four--dimensional $BF$--action~\cite{birm,horo} which
describes the
interaction between a two--form field $B^a_{\mu\nu}$ and a gauge field
$A^a_\mu$, and
\eq
 F^a_{\mu\nu}=\partial_\mu A^a_\nu - \partial_\nu A^a_\mu +
  f^{abc}A^b_\mu A^c_\nu\ .
\eqn{ftensor}

The action \equ{actioinv} has the symmetries~:
\eq
\begin{array}{lcl}
\delta^{(1)}A^a_\mu &=& -\left(\partial_\mu\theta^a+f^{abc}A^b_\mu\theta^c
                         \right)\equiv-(D_\mu\theta)^a\\
\delta^{(1)}B^a_{\mu\nu} &=& -f^{abc}B^b_{\mu\nu}c^c
\end{array}
\eqn{delta1}
and
\eq
\begin{array}{lcl}
\delta^{(2)}A^a_\mu &=& 0\\
\delta^{(2)}B^a_{\mu\nu} &=& -\left( (D_\mu\varphi_\nu)^a -
                        (D_\nu\varphi_\mu)^a\right)\ ,
\end{array}
\eqn{delta2}
with $\theta^a$ and $\varphi^a$ local parameters.

The corresponding gauge--fixing action, in a Landau--type gauge,
 reads~\cite{gms}~:
\eq\ba{rl}
S_{gf} =  \intx  \LP &\!\!
      b^{a}\partial A^{a} + {\bar c}^{a} \partial^{\mu} {(D_{\mu}c)}^{a} +
       h^{a\nu}(\partial^{\mu}B^{a}_{\mu\nu}) + \omega^{a}\partial\xi^{a} +
       h^{a}_{\mu}(\partial^{\mu}e^{a})  \\
 &\!\! + \omega^{a}\lambda^{a} +
       (\partial_{\mu} {\bar \xi}^{a\mu})\lambda^{a} -
       (\partial^{\mu}{\bar \phi}^{a})
     \left[{\ }{(D_{\mu}\phi)^{a}}  + f^{abc}c^{b}\xi_{\mu}^{c}{\ }\right] \\
 &\!\! - (\partial^{\mu} {\bar \xi}^{a\nu})
         \left[{\ }{(D_{\mu}\xi_{\nu})^{a}} - {(D_{\nu}\xi_{\mu})^{a}} +
          f^{abc}B^{b}_{\mu\nu}c^{c} {\ }\right] \\
 &\!\! +{1 \over 2} f^{abc} \varepsilon^{\mu\nu\rho\sigma}
         (\partial_{\mu}{\bar \xi}^{a}_{\nu})
         (\partial_{\rho}{\bar \xi}^{b}_{\sigma})\phi^{c} {\ }{\ }\RP
\ea\eqn{gaugeaction}
where $(c,\bar{c},b)$, $(\xi,\bar\xi, h)$ are the
ghosts, the antighosts and the Lagrangian multipliers for the
transformations \equ{delta1} and \equ{delta2}, while the fields
$(\phi,\bar\phi,\omega)$
take into account a further degeneracy due to the well--known~\cite{grisaru}
existence of zero modes in the transformations \equ{delta2}.

All the fields belong to the adjoint representation of the gauge group $G$,
assumed to be simple. The dimensions and the
Faddeev--Popov charges of the fields are

\begin{center}
\begin{tabular}{|l|r|r|r|r|r|r|r|r|r|r|r|r|r|}\hline
&$A$ & $B$ & $c$ & $\bar{c}$&$b$& $\xi$ &
$\bar\xi$& $h$&$\phi$&$\bar\phi$&$\omega$&$e$&$\lambda$
\\ \hline
dim&1&2&0&2&2&1&1&1&0&2&2&2&2\\ \hline
$\Phi\Pi$&0&0&1&-1&0&1&-1&0&2&-2&-1&0&1\\ \hline
\end{tabular}

\vspace{.2cm}{\footnotesize
{\bf Table 1.}Dimensions and Faddeev--Popov charges of the quantum fields.}
\end{center}

The gauge--fixed action $(S_{inv}+S_{gf})$ turns out
to be invariant under the two following sets of transformations~~\cite{gms}:

\eq\ba{l}
s A^{a}_{\mu}{\ } = - {(D_{\mu}c)}^{a}   \\
s c^{a}{\ }{\ } = {1 \over 2}f^{abc}c^b c^c  \\
s \xi^{a}_{\mu}{\ }{\ } =  {(D_{\mu}\phi)}^{a} +
                         f^{abc}c^{b}\xi_{\mu}^{c}    \\
s \phi^{a}{\ }{\ }   = f^{abc}c^{b}\phi^{c}       \\
s B^{a}_{\mu\nu} = - {(D_{\mu}\xi_{\nu} - D_{\nu}\xi_{\mu})}^{a}
                 - f^{abc}B^{b}_{\mu\nu}c^{c}
                 + f^{abc}\varepsilon_{\mu\nu\rho\sigma}
                   (\partial^{\rho}{\bar \xi}^{b\sigma})\phi^{c}    \\
s {\bar \xi}^{a}_{\mu}{\ }{\ } = h^{a}_{\mu}\ ,
            {\ }{\ }{\ }{\ }{\ }{\ }{\ }s h^{a}_{\mu}=0  \\
s {\bar c}^{a}{\ }{\ } = b^{a}{\ }\ ,
            {\ }{\ }{\ }{\ }{\ }{\ }{\ }s b^{a}=0  \\
s {\bar \phi}^{a}{\ }{\ } = \omega^{a}\ ,
            {\ }{\ }{\ }{\ }{\ }{\ }{\ }s \omega^{a}=0  \\
s e^{a}{\ }{\ } = \lambda^{a}\ ,
            {\ }{\ }{\ }{\ }{\ }{\ }{\ }{\ }s \lambda^{a}=0
                              \ ,
\ea\eqn{brs}
and
\eq\ba{l}
\delta_{\mu} A^{a}_{\nu} {\ } =
    -\varepsilon_{\nu\mu\tau\rho} \partial^{\tau} {\bar \xi}^{a\rho}  \\
\delta_{\mu} c^{a}{\ }{\ }   = - A_{\mu}^{a}            \\
\delta_{\mu} {\bar c}^{a}{\ }{\ }= 0                    \\
\delta_{\mu} b^{a}{\ }{\ } = \partial_{\mu} {\bar c}^{a}    \\
\delta_{\mu} B^{a}_{\nu\rho} =
    -\varepsilon_{\nu\rho\mu\tau} \partial^{\tau} {\bar c}^{a}   \\
\delta_{\mu} \xi^{a}_{\nu}{\ }{\ }= - B^{a}_{\mu\nu}  \\
\delta_{\mu} {\bar \xi}^{a\tau}{\ }=
                - \delta^{\tau}_{\mu} {\bar \phi}^{a}            \\
\delta_{\mu} h^{a\nu}{\ }=
               (\partial_{\mu} {\bar \xi}^{a\nu}){\ }+{\ }
               \delta^{\nu}_{\mu} \omega^{a}                 \\
\delta_{\mu} \phi^{a}{\ }{\ }= \xi^{a}_{\mu}      \\
\delta_{\mu} {\bar \phi}^{a}{\ }{\ }= 0           \\
\delta_{\mu} \omega^{a}{\ }{\ }=
               \partial_{\mu}{\bar \phi}^{a}                     \\
\delta_{\mu} \lambda^{a}{\ }{\ }=
               \partial_{\mu} e^{a}                              \\
\delta_{\mu} e^{a}{\ }{\ }= 0
                                            \ ,
\ea\eqn{susy}
with
\eq
s(S_{inv}+S_{gf})=\delta_\mu (S_{inv}+S_{gf})=0\ .
\eqn{invariance}

It is easy to verify that~:
\eq\ba{l}
\{\ s,s\ \} = 0\ +\ {equations\ of\ motion}  \\
\{\ s,\delta_\mu\ \} = \partial_\mu\ +\ {equations\  of\  motion}\\
\{\ \delta_\mu,\delta_\nu\ \} =0\ ,
\ea\eqn{onshellalg}
from which one sees that the BRS transformations \equ{brs}
are nilpotent on--shell.

Finally, the last term $S_{ext}$ in \equ{actionclass}
describes the coupling of the external sources
$\Lp \Omega, L, \gamma, D, \rho \Rp$
with the nonlinear transformations in \equ{brs}~:
\eq\ba{rl}
S_{ext} =   \intx \Lp &\!\!
        \Omega^{a\mu}( s A^{a}_{\mu} ) + L^{a}( s c^{a} ) +
\gamma^{a\mu\nu}( s B^{a}_{\mu\nu} ) +
        D^{a}( s \phi^{a} )  \\
 &\!\! + \rho^{a\mu} ( s\xi^{a}_{\mu})  +
     {1 \over 2}{\ }f^{abc} \varepsilon_{\mu\nu\rho\sigma}
     \gamma^{a\mu\nu}\gamma^{b\rho\sigma}\phi^{c} {\ }{\ }\Rp \ .
\ea\eqn{sext}

The dimensions and the Faddeev--Popov charges of these sources are~:
\begin{center}
\begin{tabular}{|l|r|r|r|r|r|}\hline
&$\Omega$&$L$&$\gamma$&$D$&$\rho$\\ \hline
dim&3&4&2&4&3\\ \hline
$\Phi\Pi$&-1&-2&-1&-3&-2\\ \hline
\end{tabular}

\vspace{.2cm}{\footnotesize {\bf Table 2.}
Dimensions and Faddeev--Popov
charges of the external fields.}
\end{center}

The complete action $\Sigma$ obeys to the following Slavnov identity~:
\eq
{\cal S}(\Sigma) = 0                           \ ,
\eqn{slavnov}
where
\eq\ba{rl}
{\cal S}(\Sigma)  =   \intx \LP &\!\!
       \fud{\Sigma}{\Omega^{a\mu}} \fud{\Sigma}{A^{a}_{\mu}}
   +
       \fud{\Sigma}{L^{a}} \fud{\Sigma}{c^{a}}
   +
       \frac{1}{2}
       \fud{\Sigma}{\gamma^{a\mu\nu}} \fud{\Sigma}{B^{a}_{\mu\nu}}
   +
       \fud{\Sigma}{D^{a}} \fud{\Sigma}{\phi^{a}}   \\
 &\!\!
   +   \fud{\Sigma}{\rho^{a\mu}} \fud{\Sigma}{\xi^{a}_{\mu}}
   +    h^{a}_{\mu} \fud{\Sigma}{{\bar \xi}^{a}_{\mu}}
   +    b^{a} \fud{\Sigma}{{\bar c}^{a}}
   +    \omega^{a} \fud{\Sigma}{{\bar \phi}^{a}}
   +    \lambda^{a} \fud{\Sigma}{e^{a}}  {\ }{\ }\RP
                                            \ .
\ea\eqn{slavident}

The corresponding linearized operator
\eq\ba{rl}
B_{\Sigma} = \intx \LP &\!\!
      \fud{\Sigma}{\Omega^{a\mu}} \fud{\ }{A^{a}_{\mu}}
  +
      \fud{\Sigma}{A^{a}_{\mu}} \fud{\ }{\Omega^{a\mu}}
  +   \fud{\Sigma}{L^{a}} \fud{\ }{c^{a}}
  +   \fud{\Sigma}{c^{a}} \fud{\ }{L^{a}}
  +   \frac{1}{2}
      \fud{\Sigma}{\gamma^{a\mu\nu}} \fud{\ }{B^{a}_{\mu\nu}}  \\
 &\!\!
  +   \frac{1}{2}
      \fud{\Sigma}{B^{a}_{\mu\nu}} \fud{\ }{\gamma^{a\mu\nu}}
  +   \fud{\Sigma}{D^{a}} \fud{\ }{\phi^{a}}
  +   \fud{\Sigma}{\phi^{a}} \fud{\ }{D^{a}}
  +   \fud{\Sigma}{\rho^{a\mu}}
      \fud{\ }{\xi^{a}_{\mu}}  \\
 &\!\!
  +   \fud{\Sigma}{\xi^{a}_{\mu}} \fud{\ }{\rho^{a\mu}}
  +    h^{a}_{\mu} \fud{\ }{{\bar \xi}^{a}_{\mu} }
  +    b^{a} \fud{\ }{{\bar c}^{a} }
  +    \omega^{a}\fud{\ }{ {\bar \phi}^{a} }
  +    \lambda^{a} \fud{\ }{e^{a} }
                            {\ }\RP  \ ,
\ea\eqn{linsl}
turns out to be nilpotent, {\it i.e.}~:
\eq
B_{\Sigma}B_{\Sigma}= 0       \ .
\eqn{nilpotency}

One has to remark that the nilpotency of $B_\Sigma$ is insured by the quadratic
term in the external source $\gamma^{a\mu\nu}$ in \equ{sext}.
As it is well known~\cite{bv},
the introduction of this term is necessary when the BRS transformations are
nilpotent on--shell and, in practice, is the only way which allows to define
an off--shell nilpotent linearized operator.

The $\delta_\mu$ invariance of $(S_{inv}+S_{gf})$
translates into a Ward identity for $\Sigma$~\cite{gms}:
\eq
{\cal W}_\mu\Sigma = \Delta^{cl}_\mu\ ,
\eqn{Widentity}
where
\eq\ba{rl}
{\cal W}_{\mu}  =  \intx \LP &\!\!
-\varepsilon_{\nu\mu\tau\rho}
    (\gamma^{a\tau\rho}+ \partial^{\tau}{\bar \xi}^{a\rho})
    \fud{\ }{A^{a}_{\nu}}
     - A^{a}_{\mu} \fud{\ }{c^{a}}
     -\frac{1}{2}\varepsilon_{\nu\rho\mu\tau}
    (\Omega^{a\tau} + \partial^{\tau}{\bar c}^{a})
    \fud{\ }{B^{a}_{\nu\rho}}  \\
 &\!\!
    - B^{a}_{\mu\nu} \fud{\ }{\xi^{a}_{\nu}}
    + (\partial_{\mu}{\bar c}^{a}) \fud{\ }{b^{a}}
    - {\bar \phi}^{a} \fud{\ }{{\bar \xi}^{a\mu}}
    + (\partial_{\mu}{\bar \xi}^{a\nu} + \delta^{\nu}_{\mu}\omega^{a})
      \fud{\ }{h^{a\nu}} \\
 &\!\!
    + \xi^{a}_{\mu} \fud{\ }{\phi^{a}}
    + (\partial_{\mu}{\bar \phi}^{a}) \fud{\ }{\omega^{a}}
    + (\partial_{\mu}e^{a}) \fud{\ }{\lambda^{a}}
    - D^{a} \fud{\ }{\rho^{a\mu}} \\
  &\!\!
    - L^{a} \fud{\ }{\Omega^{a\mu}}
    - {1 \over 4} ( \delta^{\sigma}_{\mu} \rho^{a\tau} -
                    \delta^{\tau}_{\mu} \rho^{a\sigma} )
             \fud{\ }{\gamma^{a\sigma\tau}}{\ }\RP
            \ ,
\ea\eqn{Wexpress}
and
\eq\ba{rl}
\Delta^{cl}_{\mu} = \intx \LP &\!\!
- \gamma^{a\rho\sigma}(\partial_{\mu}B^{a}_{\rho\sigma})
     - \Omega^{a\tau}(\partial_{\mu}A^{a}_{\tau})
     + L^{a}(\partial_{\mu}c^{a}) - D^{a}(\partial_{\mu} \phi^{a}) \\
 &\!\!
     + \rho^{a\tau}(\partial_{\mu}\xi^{a}_{\tau})
     + \varepsilon_{\mu\rho\sigma\nu}\Omega^{a\rho}
       (\partial^{\sigma}h^{a\nu})
     - \varepsilon_{\mu\rho\sigma\nu}\gamma^{a\rho\sigma}
       (\partial^{\nu}b^{a}) {\ }\RP
            \ .
\ea\eqn{Wbreaking}
Notice that $\Delta^{cl}_\mu$, being linear in the quantum fields, is
a classical breaking.

The ghost equation, usually valid in the Landau gauge~\cite{bps,ps}, in the
present case reads~:
\eq
{\cal G}^{a} \Sigma = \Delta^{a}\ ,
\eqn{ghostcl}
where
\eq
{\cal G}^{a} = \intx {\ }\left({\ }
    { \delta {\ } \over \delta \phi^{a} }
   -f^{abc}{\bar \phi}^{b}{ \delta {\ } \over \delta b^{c} } {\ }\right)
                                           \ ,
\eqn{ghostexpr}
and
\eq
\Delta^{a}= \intx {\ }f^{abc}\left({\ }
   {1 \over 2}
   \varepsilon_{\mu\nu\rho\sigma}\gamma^{b\mu\nu}\gamma^{c\rho\sigma}
 + \varepsilon_{\mu\nu\rho\sigma}\gamma^{b\mu\nu}
   (\partial^{\rho}{\bar \xi}^{c\sigma})
 + D^{b}c^{c} + \rho^{b\mu}A^{c}_{\mu}      {\ }\right)
                                          \ .
\eqn{ghostbreaking}

Anticommuting the ghost equation \equ{ghostcl} with the Slavnov
identity \equ{slavnov}, one finds
a further constraint, again linearly broken~:
\eq
{\cal F}^{a} \Sigma = \Xi^{a}\ ,
\eqn{fcl}
where
\eq\ba{rl}
{\cal F}^{a} = \intx {\ }f^{abc}\LP &\!\!
    -\varepsilon_{\mu\nu\rho\sigma}
     (\gamma^{b\mu\nu} + \partial^{\mu}{\bar \xi}^{b\nu})
     \fud{\ }{B^{c}_{\rho\sigma}}
     + \rho^{b\mu} \fud{\ }{\Omega^{c\mu}}
     - D^{b} \fud{\ }{L^{c}} \\
 &\!\!
     - c^{b} \fud{\ }{\phi^{c}}
     - A^{b}_{\mu} \fud{\ }{\xi^{c}_{\mu}}
     - {\bar \phi}^{b} \fud{\ }{{\bar c}^{c}}
     + \omega^{b} \fud{\ }{b^{c}} \RP
                           \ ,
\ea\eqn{exprf}
and
\eq
\Xi^{a}= \intx {\ }f^{abc}\varepsilon_{\mu\nu\rho\sigma}
              \gamma^{b\mu\nu}\partial^{\rho}h^{c\sigma}   \ .
\eqn{fbreaking}

The gauge--fixing conditions are~:
\eqa
\fud{\S}{b^a} &=& \partial A^a\nonumber\\
\fud{\S}{h^{a\nu}} &=& \partial^\mu B^a_{\mu\nu}+\partial_\nu e^a\nonumber\\
\fud{\S}{\omega^a} &=& \partial \xi^a+\lambda^a\label{gaugefixing}\\
\fud{\S}{\lambda^a} &=& -\partial \bar\xi^a-\omega^a\ ,\nonumber
\eea

As usual~\cite{piguetrouet}, commuting \equ{gaugefixing} with the Slavnov
identity \equ{slavnov}, one gets the antighost equations~:
\eqa
\partial^\mu\fud{\S}{\Omega^{a\mu}} + \fud{\S}{\bar{c}^a}&=& 0
                                                        \nonumber\\
\partial^\mu\fud{\S}{\rho^{a\mu}} - \fud{\S}{\bar\phi^a}&=& 0\nonumber\\
\fud{\S}{e^a} +\partial h^a&=& 0\label{antighosteq}\\
\partial^\mu\fud{\S}{\gamma^{a\mu\nu}} + \fud{\S}{\bar\xi^{a\nu}}
+\partial_\nu\lambda^a&=&0\ ,\nonumber
\eea
To summarize, the classical action $\Sigma$ \equ{actionclass}
is characterized by~:

$i)$  the Slavnov identity
\eq
{\cal S}(\Sigma)=0\ ;
\eqn{slavnov1}

 $ii)$ the vectorial supersymmetry
\eq
{\cal W}_\mu\Sigma = \Delta^{cl}_\mu\ ;
\eqn{susyward}

 $iii)$  the ghost and the ${\cal F}^a$--equations
\eq
{\cal G}^a\Sigma = \Delta^a
\eqn{ghosteq}
\eq
{\cal F}^a\Sigma = \Xi^a\ ;
\eqn{f}

 $iv)$ the gauge--fixing conditions \equ{gaugefixing}\ .

The operators in \equ{slavnov1}--\equ{f}
form a nonlinear algebra whose relevant part
takes the form~:
\eq\ba{l}
 B_\gamma {\cal S}(\gamma) =0 \\
 \{{\cal W}_\mu,{\cal W}_\nu\}=0 \\
 {\cal W}_\mu{\cal S}(\gamma) + B_\gamma({\cal W}_\mu\gamma-\Delta^{cl}_\mu) =
{\cal P}_\mu \gamma   \ ,
\ea\eqn{subalgebra}

\eq\ba{l}
 {\cal G}^a{\cal S}(\gamma)-
 B_\gamma({\cal G}^a\gamma-\Delta^a)={\cal F}^a\gamma-\Xi^a \\
 {\cal G}^a({\cal W}_\mu\gamma-\Delta^{cl}_\mu)-
 {\cal W}_\mu({\cal G}^a\gamma-\Delta^a)= 0 \\
 {\cal F}^a({\cal W}_\mu\gamma-\Delta^{cl}_\mu)+
 {\cal W}_\mu({\cal F}^a\gamma-\Xi^a)= 0 \\
 {\cal G}^a({\cal G}^b\gamma-\Delta^b)-
 {\cal G}^b({\cal G}^a\gamma-\Delta^a)=0 \\
 {\cal F}^a({\cal F}^b\gamma-\Xi^b)+
 {\cal F}^b({\cal F}^a\gamma-\Xi^a)=0 \\
 {\cal F}^a{\cal S}(\gamma)+
 B_\gamma({\cal F}^a\gamma-\Xi^a)= 0 \\
 {\cal G}^a({\cal F}^b\gamma-\Xi^b)-
 {\cal F}^b({\cal G}^a\gamma-\Delta^a)=0 \ ,
\ea\eqn{algebra}
where ${\cal P}_\mu$ is the translation operator
\eq
{\cal P}_\mu = \sum_{(all{\ }fields{\ }\varphi)}
        \intx {\ }(\partial_\mu\varphi)
        {\delta {\ }\over \delta \varphi}   \ ,
\eqn{translations}
and $\gamma$ is a generic functional with even Faddeev--Popov charge.

Notice that the subalgebra \equ{subalgebra} formed
by the Slavnov and by the operator
${\cal W}_\mu$ closes on the translations, which allows a supersymmetric
interpretation of the model. This feature is shared by a large class
of topological models as, for instance, the three--dimensional Chern--Simons
theory~\cite{chern} and the three--dimensional Einstein--Hilbert
gravity~\cite{eno}.

Finally, the rigid gauge invariance of the classical action is expressed by
\eq
{\cal H}^a_{rig}\Sigma =0\ ,
\eqn{rigid}
where
\eq
{\cal H}^a_{rig}=\sum_{(all{\ }fields{\ }\varphi)}\intx\
f^{abc}\varphi^b\frac{\delta}{\delta\varphi^c}\ .
\eqn{rigidexpr}

\section{Absence of counterterms}

This section is devoted to the algebraic characterization of the possible
local counterterms which are compatible with the symmetries and the constraints
\equ{gaugefixing}--\equ{f} satisfied by the classical action $\Sigma$.

We look then at the most general integrated local polynomial in the fields
$\widetilde\Sigma^{(c)}$ with dimensions four and zero ghost number
which satisfies the following stability
conditions~\cite{ms,bps,ps,piguetrouet}~:
\eq
\fud{\widetilde\Sigma^{(c)}}{b^a}
=
\fud{\widetilde\Sigma^{(c)}}{h^{a\nu}}
=
\fud{\widetilde\Sigma^{(c)}}{\omega^a}
=
\fud{\widetilde\Sigma^{(c)}}{\lambda^a}
= 0\ ,
\eqn{mult}

\eqa
\partial^\mu\frac{\delta\widetilde\Sigma^{(c)}}{\delta \Omega^{a\mu}}
         + \frac{\delta\widetilde\Sigma^{(c)}}{\delta\bar{c}^a}&=&
0\nonumber\\
\partial^\mu\frac{\delta\widetilde\Sigma^{(c)}}{\delta \rho^{a\mu}}
- \frac{\delta\widetilde\Sigma^{(c)}}{\delta\bar\phi^a}&=& 0\nonumber\\
\frac{\delta\widetilde\Sigma^{(c)}}{\delta e^a}  &=&0 \label{anti}\\
\partial^\mu\frac{\delta\widetilde\Sigma^{(c)}}{\delta \gamma^{a\mu\nu}}
+ \frac{\delta\widetilde\Sigma^{(c)}}{\delta\bar\xi^{a\nu}} &=&0\ ,\nonumber
\eea

\eq
{\cal G}^a\widetilde\Sigma^{(c)} = 0\ ,
\eqn{g}

\eq
{\cal F}^a\widetilde\Sigma^{(c)} =
{\cal H}^a_{rig}\widetilde\Sigma^{(c)}=0\ ,
\eqn{fh}

\eq
{\cal W}_\mu\widetilde\Sigma^{(c)} = 0\ ,
\eqn{wm}

\eq
B_\Sigma\widetilde\Sigma^{(c)} =0
\eqn{bsigma}

where $B_\Sigma$ is the linearized Slavnov operator defined in \equ{linsl}.

The conditions \equ{mult} and \equ{anti} imply that $\widetilde\Sigma^{(c)}$
does not depend on the fields $\lp b, h, \omega,
\lambda, e\rp$
and that the external  sources $\lp \Omega, \rho,
\gamma\rp$ and the antighosts $\lp \bar{c}, \bar\phi, \bar\xi\rp$
appear only in the combinations
$\lp\widehat\Omega, \widehat\rho,\widehat\gamma\rp$~:
\eq\ba{lcl}
{\widehat \Omega}^{\mu}{\ }  &=& \Omega^{\mu} + \partial^{\mu} {\bar c}
\\
{\widehat \gamma}^{\mu\nu}   &=& \gamma^{\mu\nu}
          + {1 \over 2}(\partial^{\mu}{\bar \xi}^{\nu} -
                        \partial^{\nu}{\bar \xi}^{\mu}) \\
{\widehat \rho}^{\mu}{\ }  &=& \rho^{\mu} - \partial^{\mu} {\bar \phi}
                           \ ,
\ea\eqn{combinazioni}
{\it i.e.}
\eq
\widetilde\Sigma^{(c)}=\widetilde\Sigma^{(c)}
 (A,c,B,\xi,\phi,\hat\gamma,\hat\Omega,L,\hat\rho,D)\ .
\eqn{contro}

{}From the ghost equation \equ{g} it follows that $\widetilde\Sigma^{(c)}$
depends only
on the space--time derivatives of the ghost $\phi$, so that
$\widetilde\Sigma^{(c)}$ can be parametrized as
\eq
\widetilde\Sigma^{(c)}=\widetilde\Sigma^{(2)}+\widetilde\Sigma^{(3)}
+\widetilde\Sigma^{(4)}\ ,
\eqn{count}
where, according to the number of fields,

\eq\ba{rl}
\widetilde\Sigma^{(2)} = \intx\LP&\!\!
                                a_1 B^a_{\mu\nu}B^{a\mu\nu} +
        a_2 \varepsilon^{\mu\nu\rho\sigma} B^a_{\mu\nu}B^a_{\rho\sigma} +
                 a_3 B^{a\mu\nu}\partial_\mu A^a_\nu \\
&\!\!
+ a_4 \varepsilon^{\mu\nu\rho\sigma} B^a_{\mu\nu}\partial_\rho A^a_\sigma
                       +a_5 (\partial_\mu A^a_\nu)(\partial^\mu A^{a\nu}) +
                          a_6 (\partial A^a)(\partial A^a) \\
&\!\!+a_7 \hat\gamma^{a\mu\nu}\partial_\mu\xi^a_\nu
                +a_8 \varepsilon^{\mu\nu\rho\sigma}
                          \hat\gamma^a_{\mu\nu}\partial_\rho\xi^a_\sigma
+                      a_9\xi^a_\mu\hat\Omega^{a\mu} \\
&\!\!+a_{10}\hat\Omega^{a\mu}\partial_\mu c^a +
                    a_{11}\phi^a\partial\hat\rho^a\RP\ ,
\ea\eqn{count1}

\eq\ba{rl}
\widetilde\Sigma^{(3)} = \intx f^{abc}\LP
              &\!\!
    b_1 A^a_\mu A^b_\nu B^{c\mu\nu}
  + b_2 \varepsilon^{\mu\nu\rho\sigma} A^a_\mu A^b_\nu B^c_{\rho\sigma}
  + b_3(\partial_\mu A^a_\nu)A^{b\mu}A^{c\nu}\\
&\!\!
  + b_4\hat\gamma^{a\mu\nu}B^b_{\mu\nu}c^c
  + b_5\varepsilon^{\mu\nu\rho\sigma}
                             \hat\gamma^a_{\mu\nu}B^b_{\rho\sigma}c^c
  + b_6\hat\gamma^{a\mu\nu}A^b_\mu\xi^c_\nu\\
&\!\!
  + b_7\varepsilon^{\mu\nu\rho\sigma}
                        \hat\gamma^a_{\mu\nu}A^b_\rho\xi^c_\sigma
  + b_8(\partial_\mu\hat\gamma^{a\mu\nu})A^b_\nu c^c
  + b_9\varepsilon^{\mu\nu\rho\sigma}
            (\partial_\mu\hat\gamma^a_{\nu\rho})A^b_\sigma c^c\\
&\!\!
  + b_{10}\hat\gamma^{a\mu\nu}(\partial_\mu A^b_\nu) c^c
  + b_{11}\varepsilon^{\mu\nu\rho\sigma}
         \hat\gamma^a_{\mu\nu}(\partial_\rho A^b_\sigma)c^c
  + b_{12}\hat\Omega^{a\mu}A^b_\mu c^c\\
&\!\!
  + b_{13}L^ac^bc^c
  + b_{14}\hat\rho^{a\mu}\xi^b_\mu c^c
  + b_{15}\hat\rho^{a\mu}(\partial_\mu c^b)c^c \RP\ ,
\ea\eqn{count2}

\eq\ba{rl}
\widetilde\Sigma^{(4)} = \intx\LP&\!\!
c_1^{abcd} A^a_\mu A^{b\mu} A^c_\nu A^{d\nu}
+c_2^{abcd} \varepsilon^{\mu\nu\rho\sigma}
               A^a_\mu A^b_\nu A^c_\rho A^d_\sigma
+c_3^{abcd} A^a_\mu A^b_\nu \hat\gamma^{c\mu\nu} c^d\\
&\!\!+c_4^{abcd}\varepsilon^{\mu\nu\rho\sigma}
        A^a_\mu A^b_\nu \hat\gamma^c_{\rho\sigma} c^d
+c_5^{abcd} D^ac^bc^cc^d
+c_6^{abcd} \hat\rho^{a\mu}A^b_\mu c^cc^d\\
&\!\!+c_7^{abcd} \hat\gamma^{a\mu\nu}\hat\gamma^b_{\mu\nu}c^cc^d
+c_8^{abcd} \varepsilon^{\mu\nu\rho\sigma}
           \hat\gamma^a_{\mu\nu}\hat\gamma^b_{\rho\sigma}c^cc^d\RP\ ,
\ea\eqn{count3}
and $\left\{a_i,\ b_i,\ c_i^{abcd}\right\}$ are arbitrary constant
parameters.

Let us consider now the ${\cal W}_\mu$--condition \equ{wm}.

Since the operator ${\cal W}_\mu$ in \equ{Wexpress} acts linearly
on all the fields, condition \equ{wm} holds separately for each
term of the decomposition \equ{count}. This property considerably simplifies
the algebraic analysis concerning ${\cal W}_\mu$.

The final result is that the ${\cal W}_\mu$--invariance of
$\widetilde\Sigma^{(c)}$ forces all the coefficients in
\equ{count1}, \equ{count2}, \equ{count3} to vanish,
implying the absence of counterterms~:
\eq
\widetilde\Sigma^{(c)}=0\ .
\eqn{nocount}

We can therefore conclude that conditions \equ{gaugefixing}--\equ{f}
completely
identify the classical action, {\it i.e.} there is no possibility for
any local deformation.

\section{Anomalies}

The purpose of this last section is to show the absence of anomalies for the
operators entering the nonlinear algebra \equ{subalgebra} and \equ{algebra}.

This result, combined with the previous one \equ{nocount}, concerning the
absence of counterterms, completes the proof of the perturbative finiteness
of the model.

In what follows we shall adopt the strategy of collecting all the symmetries
of the gauge--fixed action $(S_{inv}+S_{gf})$ into a unique operator
by means of the introduction of new global
ghosts~\cite{claudio}.
As we shall see,
this procedure turns out to be the most convenient one when dealing with
a nonlinear algebra involving several operators.

The gauge--fixing conditions \equ{gaugefixing}, the antighost equations
\equ{antighosteq} and the rigid gauge invariance \equ{rigid} are known to be
renormalizable and will be assumed to hold for the quantum vertex
functional $\Gamma$
\eq
\Gamma=\Sigma\ +\ O(\hbar)~:
\eqn{gamma}

\eqa
\fud{\Gamma}{b^a} &=& \partial A^a\nonumber\\
\fud{\Gamma}{h^{a\nu}} &=& \partial^\mu B^a_{\mu\nu}+\partial_\nu
e^a\nonumber\\
\fud{\Gamma}{\omega^a} &=& \partial \xi^a+\lambda^a\label{moltquant}\\
\fud{\Gamma}{\lambda^a} &=& -\partial \bar\xi^a-\omega^a\ ,\nonumber
\eea

\eqa
\partial^\mu\fud{\Gamma}{\Omega^{a\mu}} + \fud{\Gamma}{\bar{c}^a}&=& 0
                                                        \nonumber\\
\partial^\mu\fud{\Gamma}{\rho^{a\mu}} - \fud{\Gamma}{\bar\phi^a}&=&
0\nonumber\\
\fud{\Gamma}{e^a} +\partial h^a&=& 0\label{antiquant}\\
\partial^\mu\fud{\Gamma}{\gamma^{a\mu\nu}} + \fud{\Gamma}{\bar\xi^{a\nu}}
+\partial_\nu\lambda^a&=&0\ ,\nonumber
\eea

\eq
{\cal H}^a_{rig}\Gamma=0\ .
\eqn{rigquant}

\subsection{The ${\cal D}$ operator}

To collect all the symmetries \equ{brs}, \equ{susy}, \equ{ghostcl} and
\equ{fcl}
of the gauge--fixed action $(S_{inv}+S_{gf})$ into a unique operator,
let us define~:
\eq
{\cal Q}\equiv s + u^a{\cal G}^a + v^a{\cal F}^a_{(0)} + \eta^\mu \delta_\mu
+\theta^\mu{\cal P}_\mu -u^a\frac{\partial}{\partial v^a} -
       \eta^\mu\frac{\partial}{\partial\theta^\mu}\ ,
\eqn{calq}
where ${\cal F}^a_{(0)}$ coincides with the operator ${\cal F}^a$
in \equ{exprf} at vanishing external sources~:
\eq
{\cal F}^{a}_{(0)} = \intx {\ }f^{abc}\left(
-\varepsilon_{\mu\nu\rho\sigma}
\partial^{\mu}{\bar \xi}^{b\nu}
        { \delta {\ } \over \delta B^{c}_{\rho\sigma} }
- c^{b}{ \delta {\ } \over \delta \phi^{c} }
- A^{b}_{\mu}{ \delta {\ } \over \delta \xi^{c}_{\mu} }
       - {\bar \phi}^{b}{ \delta {\ } \over \delta {\bar c}^{c} }
       + \omega^{b}{ \delta {\ } \over \delta b^{c} } \right)\ ,
\eqn{fzero}
and $(u^a,v^a,\eta^\mu,\theta^\mu)$ are global parameters whose quantum numbers
are~:

\begin{center}
\begin{tabular}{|l|r|r|r|r|}\hline
&$u$&$v$&$\eta$&$\theta$\\ \hline
dim&0&0&-1&-1\\ \hline
$\Phi\Pi$&3&2&2&1\\ \hline
\end{tabular}

\vspace{.2cm}{\footnotesize
{\bf Table 3.}Dimensions and $\Phi\Pi$--charges of the parameters.}
\end{center}

It is easily seen that the operator ${\cal Q}$  describes a symmetry
of the gauge--fixed action~:
\eq
{\cal Q}\left(S_{inv}+S_{gf}\right)=0\ ,
\eqn{}
and, as it happens for the BRS transformations \equ{brs},
\eq
{\cal Q}^2=0\ +\ {equations\ of\ motion}\ ,
\eqn{}
{\it i.e.} ${\cal Q}$ is nilpotent on--shell.

Introducing the modified source term
\eq\ba{rl}
S^{({\cal Q})}_{ext}=\intx\LP&\!\!
\Omega^{a\mu}( {\cal Q}A^{a}_{\mu} ) + L^{a}( {\cal Q}c^{a} ) +
\gamma^{a\mu\nu}( {\cal Q}B^{a}_{\mu\nu} ) \\
&\!\!
+ D^{a}( {\cal Q}\phi^{a} )
+ \rho^{a\mu} ( {\cal Q}\xi^{a}_{\mu})\\
&\!\!
+\varepsilon_{\mu\nu\rho\sigma}\left(
{1 \over 2}f^{abc}\gamma^{a\mu\nu}\gamma^{b\rho\sigma}\phi^{c}
+\eta^\mu\Omega^{a\nu}\gamma^{a\rho\sigma}+f^{abc}v^a\gamma^{b\mu\nu}
\gamma^{c\rho\sigma}\right)\RP\ ,
\ea\eqn{sextd}
the new total action
\eq
{\cal I}=S_{inv}+S_{gf}+S^{({\cal Q})}_{ext}
\eqn{dazione}
obeys the following generalized Slavnov identity
\eq
{\cal D}({\cal I})=0\ ,
\eqn{dslavnov}
where
\eq\ba{rl}
{\cal D}({\cal I})=\intx\LP&\!\!
\fud{{\cal I}}{\Omega^{a\mu}}
\fud{{\cal I}}{A^{a}_{\mu}}
+
\fud{{\cal I}}{L^{a}}
\fud{{\cal I}}{c^{a}}
+
\frac{1}{2}\fud{{\cal I}}{\gamma^{a\mu\nu}}
\fud{{\cal I}}{B^{a}_{\mu\nu}}
+
\fud{{\cal I}}{D^{a}}
\fud{{\cal I}}{\phi^{a}}
\\&\!\!
+
\fud{{\cal I}}{\rho^{a\mu}}
\fud{{\cal I}}{\xi^{a}_{\mu}}
+
({\cal Q}\bar{c}^a)
\fud{{\cal I}}{{\bar c}^{a}}
+
({\cal Q}b^a )
\fud{{\cal I}}{b^a}
+
({\cal Q}\bar\xi^a_\mu)
\fud{{\cal I}}{\xi^a_\mu}
\\&\!\!
+
({\cal Q}h^a_\mu)
\fud{{\cal I}}{h^a_\mu}
+
({\cal Q}\bar\phi^a)
\fud{{\cal I}}{{\bar \phi}^{a}}
+
({\cal Q}\omega^a)
\fud{{\cal I}}{\omega^a}
+
({\cal Q}e^a)
\fud{{\cal I}}{e^{a}}
\\&\!\!
+
({\cal Q}\lambda^a)
\fud{{\cal I}}{\lambda^a}\RP {\ }{\ }
-u^a\pad{{\cal I}}{v^a}
-\eta^\mu\pad{{\cal I}}{\theta^\mu}\ .
\ea\eqn{dslavnovexpr}

It is remarkable to note that the modified source term \equ{sextd} leads
to a generalized Slavnov identity \equ{dslavnov} which, in contrast to what
happens for the operators ${\cal W}_\mu$, ${\cal G}^a$ and ${\cal F}^a$ in
\equ{Widentity}, \equ{ghostcl} and \equ{fcl}, surprisingly describes an
{\it exact} symmetry of the total action ${\cal I}$

Again, due to the presence in $S^{({\cal Q})}_{ext}$ of quadratic terms
in the external sources, the linearized operator
\eq\ba{rl}
D_{{\cal I}} =\intx\LP&\!\!
\fud{{\cal I}}{\Omega^{a\mu}}
\fud{}{A^{a}_{\mu}}
+
\fud{{\cal I}}{A^{a}_{\mu}}
\fud{}{\Omega^{a\mu}}
+
\fud{{\cal I}}{L^{a}}
\fud{}{c^{a}}
+
\fud{{\cal I}}{c^{a}}
\fud{}{L^{a}}\\
&\!\!
+
\frac{1}{2}\fud{{\cal I}}{\gamma^{a\mu\nu}}
\fud{}{B^{a}_{\mu\nu}}
+
\frac{1}{2}\fud{{\cal I}}{B^{a}_{\mu\nu}}
\fud{}{\gamma^{a\mu\nu}}
+
\fud{{\cal I}}{D^{a}}
\fud{}{\phi^{a}}
+
\fud{{\cal I}}{\phi^{a}}
\fud{}{D^{a}}      \\
&\!\!
+
\fud{{\cal I}}{\rho^{a\mu}}
\fud{}{\xi^{a}_{\mu}}
+
\fud{{\cal I}}{\xi^{a}_{\mu}}
\fud{}{\rho^{a\mu}}
+
({\cal Q}\bar{c}^a)
\fud{}{{\bar c}^{a}}
+
({\cal Q}b^a)
\fud{}{b^a} \\
&\!\!
+
({\cal Q}\bar\xi^a_\mu)
\fud{}{{\bar \xi}^{a}_{\mu}}
+
({\cal Q}h^a_\mu)
\fud{}{h^a_\mu}
+
({\cal Q}\bar\phi^a)
\fud{}{{\bar \phi}^{a}}
+
({\cal Q}\omega^a)
\fud{}{\omega^a}\\
&\!\!
+
({\cal Q}e^a)
\fud{}{e^{a}}
+
({\cal Q}\lambda^a)
\fud{}{\lambda^a}\RP
-u^a\pad{}{v^a}
-\eta^\mu\pad{}{\theta^\mu}\ ,
\ea\eqn{dslavnovlin}
is nilpotent~:
\eq
D_{{\cal I}}D_{{\cal I}}=0\ .
\eqn{dnil}

It is apparent now that the Slavnov identity in \equ{dslavnov} describes the
complete nonlinear algebra \equ{subalgebra} and \equ{algebra}.

The dependence of the action ${\cal I}$ on the new ghosts
$(u^a,v^a,\eta^\mu,\theta^\mu)$ is controlled by
the following equations~:
\eq
\pad{{\cal I}}{u^a} = \Delta^a_{(u)} \ , \qquad
\pad{{\cal I}}{v^a} = \Delta^a_{(v)}
\eqn{eqghostgl1}
\eq
\pad{{\cal I}}{\eta^\mu} = \Delta^{(\eta)}_\mu \ , \qquad
\pad{{\cal I}}{\theta^\mu} =
\Delta^{(\theta)}_\mu \ ,
\eqn{eqghostgl2}
where
\eqa
\Delta^a_{(u)} &=& -\intx D^a \nonumber\\
\Delta^a_{(v)} &=& \intx f^{abc}\left(
  \varepsilon_{\mu\nu\rho\sigma}\gamma^{b\mu\nu}\partial^\rho\bar\xi^{c\sigma}
  +\rho^{b\mu}A^c_\mu+D^bc^c+\frac{1}{2}\varepsilon_{\mu\nu\rho\sigma}
  \gamma^{b\mu\nu}\gamma^{c\rho\sigma}\right)\nonumber\\
\Delta^{(\eta)}_\mu &=&\intx\LP
  \varepsilon_{\mu\nu\rho\sigma}\Omega^{a\nu}(
  \partial^\rho\bar\xi^{a\sigma}+\gamma^{a\rho\sigma})-L^aA^a_\mu-
  \varepsilon_{\mu\nu\rho\sigma}\gamma^{a\rho\sigma}\partial^\nu\bar{c}^a
\nonumber\\
&&\qquad\quad
  -\rho^{a\nu}B^a_{\mu\nu}+D^a\xi^a_\mu\RP\label{rotture}\\
\Delta^{(\theta)}_\mu &=&\intx\left(
  -\Omega^{a\nu}\partial_\mu A^a_\nu+L^a\partial_\mu c^a-\gamma^{a\rho\sigma}
  \partial_\mu
B^a_{\rho\sigma}+\rho^{a\nu}\partial_\mu\xi^a_\nu-D^a\partial_\mu
   \phi^a\right)\ ,\nonumber
\eea
which, being linear in the quantum fields, represent classical breakings.

Notice that
\eq
\Delta^a_{(v)}=\Delta^a\ ,
\eqn{}
where $\Delta^a$ is the classical breaking of the ghost equation \equ{ghostcl}.

The operators \equ{dslavnovexpr} and \equ{eqghostgl1}, \equ{eqghostgl2}
form the following nonlinear algebra~:
\eq
\pad{}{u^a}{\cal D}(\gamma) +
D_\gamma \left(\pad{\gamma}{u^a}-\Delta^a_{(u)}\right)=
{\cal G}^a\gamma-\pad{\gamma}{v^a}
\eqn{dalgebra1}
\eq\ba{lcl}
\pad{}{v^a}{\cal D}(\gamma) &-&
D_\gamma \left(\pad{\gamma}{v^a}-\Delta^a_{(v)}\right)=\\
&{\cal F}^a\gamma &
+ \intx\varepsilon_{\mu\nu\rho\sigma}f^{abc}
(\partial^\rho\gamma^{b\mu\nu})\left(h^{c\sigma}-\eta^\sigma\bar\phi^c+
\theta^\lambda\partial_\lambda\bar\xi^{c\sigma}\right)
\ea\eqn{dalgebra2}
\eq\ba{lcl}
\pad{}{\eta^\mu}{\cal D}(\gamma) &-&
D_\gamma \left(\pad{\gamma}{\eta^\mu}-
\Delta^{(\eta)}_\mu\right)= \\
&{\cal W}_\mu\gamma &  -\pad{\gamma}{\theta^\mu}
+\intx\varepsilon_{\mu\nu\rho\sigma}
(\partial^\rho\Omega^{a\nu})
\left(
h^{a\sigma}-\eta^\sigma\bar\phi^a+
\theta^\lambda\partial_\lambda\bar\xi^{a\sigma}
\right)\\
&&-\intx\varepsilon_{\mu\nu\rho\sigma}
(\partial^\nu\gamma^{a\rho\sigma})\left(b^a-f^{abc}v^b\bar\phi^c+
\theta^\lambda\partial_\lambda\bar{c}^a\right)
\ea\eqn{dalgebra3}
\eq
\pad{}{\theta^\mu}{\cal D}(\gamma) +
D_\gamma \left(\pad{\gamma}{\theta^\mu}-\Delta^{(\theta)}_\mu
\right)={\cal P}_\mu\gamma\ ,
\eqn{dalgebra4}
where $\gamma$ is a generic functional with even $\Phi\Pi$--charge.

It is important at this point to spend a few words on this
construction, the motivation for the introduction
of the operator ${\cal D}(\gamma)$
is that the search of the anomalies for each single operator of the nonlinear
algebra \equ{subalgebra}--\equ{algebra} is now reduced to the
characterization of a unique anomaly for the
Slavnov operator ${\cal D}(\gamma)$, {\it i.e.} if the quantum vertex
functional
\eq
\Gamma^{({\cal Q})}={\cal I}+O(\hbar)
\eqn{gammad}
satisfies the Slavnov identity
\eq
{\cal D}(\Gamma^{({\cal Q})})=0
\eqn{}
and the global ghost equations \equ{eqghostgl1}, \equ{eqghostgl2}
then the vertex functional
\eq
\Gamma \equiv \left. \Gamma^{({\cal Q})} \right |_{u=v=\eta=\theta=0}
\eqn{}
obeys the equations
\eq
{\cal S}(\Gamma)=0
\eqn{aim1}
and
\eq
{\cal G}^a\Gamma=\Delta^a\ ,\qquad
{\cal F}^a\Gamma=\Xi^a
\eqn{aim2}
\eq
{\cal W}_\mu\Gamma=\Delta^{cl}_\mu\ ,\qquad
{\cal P}_\mu\Gamma = 0\ ,
\eqn{aim3}
which imply the absence of anomalies for the operators
\equ{slavnov1}--\equ{f}.

Indeed, replacing $\gamma$ by $\Gamma^{({\cal Q})}$ in
\equ{dalgebra1}--\equ{dalgebra4}
one finds the identities~:
\eqa
{\cal G}^a\Gamma^{({\cal Q})} &=&
\frac{\partial\Gamma^{({\cal Q})} }{\partial v^a} = \Delta^a_{(v)}
             \equiv \Delta^a\nonumber\\
{\cal F}^a\Gamma^{({\cal Q})}  &=& - \intx\varepsilon_{\mu\nu\rho\sigma}f^{abc}
(\partial^\rho\gamma^{b\mu\nu})\left(h^{c\sigma}-\eta^\sigma\bar\phi^c+
\theta^\lambda\partial_\lambda\bar\xi^{c\sigma}\right)\nonumber\\
{\cal W}_\mu\Gamma^{({\cal Q})}  &=&
\frac{\partial\Gamma^{({\cal Q})} }{\partial \theta^\mu}
 -\intx\varepsilon_{\mu\nu\rho\sigma}
(\partial^\rho\Omega^{a\nu})
\left(
h^{a\sigma}-\eta^\sigma\bar\phi^a+
\theta^\lambda\partial_\lambda\bar\xi^{a\sigma}
\right)
\label{aimdim}\\
&&\ \ \ \ \ \ \ \ +\intx\varepsilon_{\mu\nu\rho\sigma}
(\partial^\nu\gamma^{a\rho\sigma})\left(b^a-f^{abc}v^b\bar\phi^c+
\theta^\lambda\partial_\lambda\bar{c}^a\right)\nonumber\\
{\cal P}_\mu\Gamma^{({\cal Q})}  &=& 0\ ,\nonumber
\eea
which, taken at vanishing global ghosts,
reduce exatcly to the identities \equ{aim2}, \equ{aim3}, while
\eq
\left.{\cal D}(\Gamma^{({\cal Q})})\right|_{u=v=\eta=\theta=0}
\equiv{\cal S}(\Gamma)=0\ .
\eqn{aimdim1}

\subsection{Absence of anomalies}

For what concerns the renormalization of the global ghost equations
\equ{eqghostgl1}, \equ{eqghostgl2},  it is easy to show
that they can be implemented at the quantum level,
{\it i.e.}~:
\eq
\pad{\Gamma^{({\cal Q})}}
{u^a} = \Delta^a_{(u)} \ , \qquad
\pad{\Gamma^{({\cal Q})}}{v^a} = \Delta^a_{(v)}
\eqn{gammagl1}
\eq
\pad{\Gamma^{({\cal Q})}}{\eta^\mu} = \Delta^{(\eta)}_\mu \ , \qquad
\pad{\Gamma^{({\cal Q})}}{\theta^\mu} =
\Delta^{(\theta)}_\mu \ ,
\eqn{gammagl2}

Let us now discuss the renormalization
of the Slavnov identity \equ{dslavnov}.

According to the Quantum Action Principle~\cite{lam},
the least order breaking ${\cal A}$ of the Slavnov identity
\eq
{\cal D}(\Gamma^{({\cal Q})})={\cal A} + O(\hbar{\cal A})
\eqn{}
is an integrated local functional with dimensions four and Faddeev--Popov
charge $+1$, which, at the lowest nonvanishing order in $\hbar$, satisfies
the consistency condition
\eq
D_{{\cal I}}{\cal A}=0\ .
\eqn{cohom}

To study the cohomology of the nilpotent operator $D_{{\cal I}}$, we introduce
the filtration
\eq
{\cal N} = u^a \pad{}{u^a} + v^a \pad{}{v^a} +
           \eta^\mu \pad{}{\eta^\mu} + \theta^\mu \pad{}{\theta^\mu}\ .
\eqn{filtration}

Thanks to the property \equ{aimdim1}, ${\cal N}$ decomposes the operator
$D_{{\cal I}}$ as~:
\eq
D_{{\cal I}}=D^{(0)}+D^{(R)}\ ,
\eqn{deco}
with
\eq
D^{(0)} = B_\Sigma - u^a\pad{}{v^a} - \eta^\mu\pad{}{\theta^\mu}\ ,
\eqn{dzero}
and $B_\Sigma$ is the linearized Slavnov operator defined in \equ{linsl}.
Due to the nilpotency of $B_\Sigma$, it follows that
\eq
D^{(0)}D^{(0)}=0\ .
\eqn{nilpo}

Since the cohomology of $D_{{\cal I}}$ is isomorphic to a subspace of that
of $D^{(0)}$~\cite{dixon}, we proceed to the characterization of the cohomology
of this latter operator, {\it i.e.} we study the equation
\eq
D^{(0)} X =0\ ,
\eqn{comdzero}
in the space of integrated local functionals with dimension four and
Faddeev--Popov charge $+1$.

Expression \equ{dzero} shows that the global ghosts $(u,\ v)$ and $(\eta,\
\theta)$ appear in BRS doublets and it is known~\cite{dixon} that the
cohomology cannot depend on such couples.

The equation \equ{comdzero} is then equivalent to~:
\eq
B_\Sigma \Delta =0\ ,
\eqn{combsigma}
where $\Delta$ is an integrated local functional independent from
the pairs of global ghosts which, on the other hand, are not reintroduced
by $B_\Sigma$ \equ{linsl}.

Writing $\Delta$ in terms of differential forms~:
\eq
\Delta=\int \Delta^1_4(x)\ ,
\eqn{}
the equation \equ{combsigma} can be translated into a local one as~:
\eq
B_\Sigma\Delta^{1}_{4} + d \Delta^{2}_{3} =0\ ,
\eqn{local}
where $d$ is the exterior derivative and
$\Delta^{1}_{4}$, $\Delta^{2}_{3}$ are forms of degree four
and three and ghost number one and two respectively, according to the notation
\eq
\Delta^{p}_{q} \ \ ;\ \
\left\{
\begin{array}{lcl}
p &=& \mbox{\it ghost number}\\
q &=& \mbox{\it form degree}\\
\end{array}
\right.
\eqn{formnot}

The identity \equ{local} yields a
sequence of descent equations~\cite{zumino}
\eq
B_\Sigma\Delta^{1}_{4} + d\Delta^{2}_{3}=0
\eqn{disc1}
\eq
B_\Sigma\Delta^{2}_{3} + d\Delta^{3}_{2}=0
\eqn{disc2}
\eq
B_\Sigma\Delta^{3}_{2} + d\Delta^{4}_{1}=0
\eqn{disc3}
\eq
B_\Sigma\Delta^{4}_{1} + d\Delta^{5}_{0}=0
\eqn{disc4}
\eq
B_\Sigma\Delta^{5}_{0} = 0\ ,
\eqn{disc5}
whose most general expression for the $\Delta^{5}_{0}$--cocycle is
\eq
\Delta^{5}_{0} =
a_1^{abcde}c^ac^bc^cc^dc^e + a_2^{abcd}\phi^ac^bc^cc^d +
a_3^{abc}\phi^a\phi^bc^c\ .
\eqn{delta5}

The invariance condition \equ{disc5} implies that
\eq
\Delta^{5}_{0} =
rd^{apq}f^{pbc}f^{qde}
c^ac^bc^cc^dc^e + B_\Sigma\widetilde\Delta^4_0
\eqn{soluzione}
where $d^{abc}$ is the completely symmetric tensor of rank three
and $r$ an arbitrary coefficient.

This cocycle yields the usual nonabelian
gauge anomaly~\cite{zumino} whose absence, in this case, is insured by the
fact that all the fields belong to the adjoint representation of the gauge
group $G$. It is well known indeed that the triangle anomaly generates a
symmetric tensor $d^{abc}$ with a nonvanishing coefficient only if the model
contains fields belonging to a complex representation of $G$.
This means that the coefficient $r$ in \equ{soluzione} vanishes.

We have then proven that the local cohomology of $B_\Sigma$
in the five--charged Faddeev--Popov sector is empty. This implies that the
equation \equ{disc4} reduces to a problem of a local cohomology instead of a
modulo--$d$ one, leading us to the study of the local cohomology of the
$B_\Sigma$--operator.

To do this, we introduce a further filtering operator~\cite{dixon}~:
\eq
\widetilde{\cal N}={\cal N}_A+{\cal N}_B+{\cal N}_c
+{\cal N}_\xi
+{\cal N}_{\hat\gamma}+{\cal N}_{\hat\Omega}+{\cal N}_L
+{\cal N}_D+{\cal N}_{\hat\rho}+{\cal N}_\phi
\eqn{filtro}
where ${\cal N}_\varphi$ is the counting operator~:
\eq
{\cal N}_\varphi = \intx\ \varphi\frac{\delta}{\delta\varphi}\ .
\eqn{deffiltro}

$\widetilde{\cal N}$ decomposes the operator $B_\Sigma$ as~:
\eq
B_\Sigma = B_\Sigma^{(0)} + B_\Sigma^{(R)}
\eqn{decbsigma}
where
\eq\ba{rl}
B_\Sigma^{(0)}=\intx\LP &\!\!
-\partial_\mu c^a \fud{}{A^a_\mu}
-\partial_\mu\xi^a_\nu\fud{}{B^a_{\mu\nu}}
+\partial_\mu\phi^a\fud{}{\xi^a_\mu}
-\frac{1}{4}\varepsilon^{\mu\nu\rho\sigma}\partial_\rho A^a_\sigma
        \fud{}{\hat\gamma^{a\mu\nu}}\\
&\!\!
-\frac{1}{2}\varepsilon^{\mu\nu\rho\sigma}\partial_\nu B^a_{\rho\sigma}
        \fud{}{\widehat\Omega^{a\mu}}
+2\partial_\nu\hat\gamma^{a\mu\nu}\fud{}{\hat\rho^{a\mu}}
-\partial\hat\Omega^a\fud{}{L^a}
-\partial\hat\rho^a\fud{}{D^a}\RP\ .
\ea\eqn{bzero}
and
\eq
B_\Sigma^{(0)}B_\Sigma^{(0)}=0\ .
\eqn{nilbzero}

Following~\cite{ps,dixon}, it is not difficult to show that the local
cohomology of
$B_\Sigma^{(0)}$ depends only on the undifferentiated ghosts $(c,\
\phi)$.

Since $c$ and $\phi$ are both dimensionless, it follows that
the most general solution for the higher cocycle $\Delta^1_4$ is a
$B_\Sigma^{(0)}$--coboundary modulo a total derivative, implying then the
vanishing of the cohomology of $B_\Sigma$.

Moreover, since the cohomology of $D_{{\cal I}}$ is isomorphic~\cite{dixon}
to a subspace of that of $B_\Sigma$, it follows that the general solution of
the
equation \equ{cohom} is
\eq
{\cal A}=D_{{\cal I}}\widehat{\cal A}\ .
\eqn{finalresult}

This concludes the proof of the absence of anomalies.

\vspace{2cm}
\begin{center}
\bf Acknowledgements
\end{center}
We would like to thank Alberto Blasi and Olivier Piguet for interesting
discussions. N.M. is grateful to the {\it D\'epartement
de Physique Th\'eorique de l'Universit\'e de Gen\`eve} for the kind
hospitality during the preparation of this work.

\end{document}